# A Two-Dimensional Resistor Network Model for Transition-Edge Sensors with Normal Metal Features

Daikang Yan[1,2], Lisa M. Gades[1], Tejas Guruswamy[1], Antonino Miceli[1,2], Umeshkumar M. Patel[1] and Orlando Quaranta[1,2]
[1] Argonne National Laboratory, Argonne, IL 60439 USA
[2] Northwestern University, Evanston, IL 60208 USA
Email: oquaranta@anl.gov

**Abstract**
Transition-edge sensors (TESs) can be used in high-resolution photon detection, exploiting the steep slope of the resistance in the superconducting-to-normal transition edge. Normal metal bars on the TES film are commonly used to engineer its transition shape, namely the dependence of resistance on temperature and current. This problem has been studied in one dimension, however until now, there have been no predictive models of the influence of two-dimensional (2-D) normal metal features on the TES transition shape. In this work, we approach this problem by treating the TES as a 2-D network of resistors, the values of which are based on the two-fluid model. We present a study of the behavior of devices with different 2-D geometric features. Our 2-D network model is capable of predicting how typical TES geometry parameters, such as number of bars, bar spacing, and overall dimensions, influence device behavior and thus is a powerful tool to guide the engineering of new TES devices.

## 1. Introduction

Transition-edge sensors (TESs) are typically made of low-temperature superconductor materials or multilayers, and have been widely used and studied as bolometers and calorimeters. In the small signal, linearized limit, the signal response of a TES can be expressed as a set of electrothermal differential equations and solved analytically [1]. However, the TES response and specifically resistance as a function of current and temperature $R(I, T)$ is usually nonlinear over the full superconducting transition width [2][3]. Knowledge of the full nonlinear $R(I, T)$ function is necessary to correctly describe the device response to large signals as well as its saturation and dynamic range. Multiple models have been used to describe the nonlinear transition shape of TESs [3][4], including the resistively and capacitively shunted junction (RCSJ) model [5] for small dimension devices, that treats the TES as a weak link between superconducting leads; and the two-fluid model [6][7] that describes the TES current as a superposition of a superconductive (Cooper pair) and a normal (quasiparticle) current. The latter has shown good quantitative agreement with the current-voltage (I-V) curves measured from a 350 μm × 350 μm Mo-Cu bilayer TES [7]. However, it is also known that the geometry of noise mitigating normal metal bars/banks patterned on the superconducting film influences the $R(I, T)$ transition shape [8][9]; this is not accounted for in the simple two-fluid model. The effect of normal metal bars has been studied in a one-dimensional (1-D) scheme utilizing the Usadel equation [8], but has not been addressed in two dimensions (2-D). In this work, we present a resistor network model to calculate the TES transition shape given 2-D normal metal features of arbitrary geometry. The current distributions and $R(I, T)$ surfaces for some example TESs are solved, and dependence on the normal metal features is shown.

## 2. Two-dimensional resistor network model

### 2.1. Four-terminal resistor units

In this model, the TES is represented as a 2-D "film" divided into small square units. Each unit is a square containing four identical resistors, one per side (Fig.1). In this way, currents can flow in both the longitudinal and the transverse direction. The neighboring units join at the resistor nodes. In the case of a unit representing a part of the TES where a normal metal feature is present (bars or banks - in orange in Fig. 1, labeled "normal metal"), the resistance is fixed to be $R_{metal}$, as we assume these units are completely in the normal state due to the proximity effect. In the case of a unit where no normal metal features are present (in blue in Fig. 1, labelled "transition region"), the resistance $R$ depends on both the total current $I$ flowing through the unit, and its temperature T. The net current flowing through a unit is calculated as $I = (I_x^2 + I_y^2)^{1/2}$ where $I_x$ and $I_y$ are the current components in the respective directions, while the temperature is assumed to be the same across the TES, ignoring the Joule heating effect and any potential non-uniform cooling in the film. The $R(I, T)$ relation for each transition region unit is defined by the two-fluid model and will be introduced in Section 2.2.

The resistor network must obey a matrix equation representing Ohm's Law

$$[S] \cdot \vec{V}_{node} = \vec{I}_{node}. \qquad (1)$$

For an $m \times n$ 2-D network, $\vec{V}_{node}$ is an $(m+1)(n+1)$ column vector that consists of the node voltages, while $\vec{I}_{node}$ is an $(m+1)(n+1)$ column vector that consists of the net node currents. $[S]$ is an $(m+1)(n+1) \times (m+1)(n+1)$ matrix consisting of the electrical conductances between each node and is constructed based on the values of the resistors in the network. By Kirchhoff's current law, the net current at each node must be zero, except for the two nodes representing the contacts with the bias leads, where the net current is $\pm I_{bias}$. Here the sign defines whether the current is flowing in or out of the node (Fig. 1).

Table I: The geometry and model parameters of the seven TESs simulated

| geometry | 1/2-sq. 3-bar | 1/2-sq. 2-bar | 1-sq. 4-bar | 1-sq. 3-bar | 2-sq. 5-bar | 2-sq. 4-bar | 2-sq. 3-bar |
|---|---|---|---|---|---|---|---|
| Number of units between bars | 4 | 5 | 4 | 6 | 5 | 7 | 9 |
| $T_c$ [mK] | 71.3 | 73.5 | 72.8 | 73.4 | 73.0 | 73.9 | 76.3 |
| $c_I$ | 0.79 | | | | | | |
| $c_R$ | 0.25, 0.5, 1 | | | | | | |
| $G$ [pW/K] | 100 | | | | | | |
| $T_{bath}$ [mK] | 55 | | | | | | |

Matrix [$S$] is singular, and is of order $(m+1)(n+1)\times(m+1)(n+1)-1$. Physically, this is because the node voltages are relative to an arbitrary ground voltage. This is resolved by fixing the voltage of one node. Here we set the voltage of the node connected to the ground to be zero (although it can be any arbitrary value). This changes the ground node Kirchhoff equation from

$$\vec{S} \cdot \vec{V}_{node} = -I_{bias}, \quad (2a)$$

to
$$[0,0,\dots,1,\dots 0] \cdot \vec{V}_{node} = 0. \quad (2b)$$

where the $\vec{S}$ vector is all zero except the value one at the position that multiplies with $V_{ground}$. The modified Kirchhoff equation set is now linearly independent, therefore the node voltages can be solved by

$$\vec{V}_{node} = [\,S\,]^{-1}\vec{I}_{node}. \quad (3)$$

The current between any two nodes can then be calculated by dividing the voltage difference by the connecting resistance.

Fig. 2 illustrates a simple network consisting of two four-terminal units. Following the steps introduced above, its node voltages can be calculated as Eq. (4).

The geometry of the TES determines the 2-D network and the matrix [$S$]. Figure 1 shows the network model of a 150 μm × 150 μm TES film with parallel normal metal banks and three perpendicular normal metal bars. The width of the banks and bars is ~ 10 μm, therefore the TES is divided into 5 μm × 5 μm squares, providing sufficient resolution to accurately represent the geometry. The units representing the normal metal region use resistor values of $R_{metal}$ = 8.94 mΩ, and the transition region units use a normal state resistance of $R_n$ = 23.54 mΩ in the two-fluid model calculation. These numbers are based on Ref. [9], but with the four-terminal unit configuration the resistor values are doubled relative to the measured sheet resistance, to ensure that any square section of the overall network has the correct total resistance. Theoretically, the resistance of the superconducting leads is zero. However, to avoid null values in the simulation, a very small value of 1 nΩ is used instead.

2.2. Resistance-current relation

The two-fluid model defines the resistance of a superconductor in the transition region $R(I, T)$ as:

$$\begin{bmatrix} V_1 \\ V_2 \\ V_3 \\ V_4 \\ V_5 \\ V_6 \end{bmatrix} = \begin{bmatrix} \frac{2}{R_a} & \frac{-1}{R_a} & 0 & \frac{-1}{R_a} & 0 & 0 \\ \frac{-1}{R_a} & (\frac{2}{R_a}+\frac{2}{R_b}) & -\frac{1}{R_b} & 0 & (\frac{-1}{R_a}+\frac{-1}{R_b}) & 0 \\ 0 & \frac{-1}{R_b} & \frac{2}{R_b} & 0 & 0 & \frac{-1}{R_b} \\ \frac{-1}{R_a} & 0 & 0 & \frac{2}{R_b} & \frac{-1}{R_a} & 0 \\ 0 & (\frac{-1}{R_a}+\frac{-1}{R_b}) & 0 & \frac{-1}{R_a} & (\frac{2}{R_a}+\frac{2}{R_b}) & \frac{-1}{R_b} \\ 0 & 0 & 0 & 0 & 0 & 1 \end{bmatrix} \cdot \begin{bmatrix} 0 \\ 0 \\ 0 \\ I_{bias} \\ 0 \\ 0 \end{bmatrix}. \quad (4)$$

$$R(I,T) = \begin{cases} \left[1 - \dfrac{c_I I_{c0}}{I}\left(1 - \dfrac{T}{T_c}\right)^{3/2}\right] c_R R_n, & I > c_I I_{c0}\left(1 - \dfrac{T}{T_c}\right)^{3/2}, \quad (5a) \\ 0, & I < c_I I_{c0}\left(1 - \dfrac{T}{T_c}\right)^{3/2}. \quad (5b) \end{cases}$$

This model has shown good agreement with experimental results when applied to the entire TES [4][7]. In this equation, $T_c$ is the critical temperature, $I_{c0}$ is the critical current at zero temperature, and $c_I$ and $c_R$ are unitless coefficients. $T$ is the temperature, assumed to be the same for all the units, and I is the current passing through each unit. In the case of our 2-D network model, the current $I$ depends on the distribution of $R$.

Because $R$ and $I$ are mutually dependent, and because this relationship is nonlinear, as suggested by Eq. (5a), the solution to Eq. (3) can only be obtained by numerical methods. We choose to solve Eq. (3) iteratively with Newton-Krylov method, using finite-differences to estimate function derivatives. To avoid non-convergence of the numerical solver in the zero-derivative region when $I < I_{c0}$ in Eq. (5b), $R(I, T)$ is modified to have a constant slope as a function of current below 5% $R_n$, as shown in Fig. 3. In order to evaluate the influence of this modification, we simulated via the 2-D resistor network the $R(I, T)$ of a TES with no normal metal structures, and compared it to the prediction of the two-fluid model applied to an equivalent single-body TES. The two results should be effectively identical, with a uniform current distribution through the 2-D network, resulting from a uniformly segmented resistance. The only difference between the two approaches arises from the modification in the 2-D network model of the two-fluid $R(I, T)$ below 5% $R_n$. The comparison shows that the difference caused by this modification is below 5% $R_n^{\text{total}}$ ($R_n^{\text{total}}$ is the total normal resistance of the device), which we deem negligible. With this point verified, we believe this 2-D network model is capable of describing how typical TES geometry parameters, such as number of bars, bar spacing, and overall dimensions, influence device behavior. The results of these simulations for some specific TES designs, chosen to be similar to those of Ref. [10], are described in the following section.

**3. Results and discussion**

3.1. Simulation parameters

To explore the influence of some common TES design parameters, we decided to focus on a set of seven different designs labeled with the following scheme: *x*-sq. *y*-bars, where *x* represents the aspect ratio (width/height) of the device and *y* represents the number of bars present. It should be expected that changing the width of the device results in a different total TES resistance, and therefore a different trajectory on the $R(I, T)$ surface under voltage bias, while changing the number of bars may affect the current distribution in the device, defining a different shape for the $R(I, T)$ surface. The specific designs under examination are based on those described in Ref. [10] and illustrated in Fig. 4. We also used the material and device parameters measured in Ref. [10] (see Table I), providing means to validate the model against the experimental measurements presented in that work.

However, not all the parameters required here have an immediate equivalent in that paper – in particular, the measured two-fluid model parameters of Ref. [10] are those appropriate to an entire TES including bilayer and normal metal features, while in the 2-D resistor network both units with and without normal metal features are present at the same time, therefore some assumptions had to be made and are described below.

In Table I, the critical temperature of a unit resistor has been defined as dependent on the overall design. This approximates the lateral proximity effect induced by the normal metal features, as our model does not directly account for this effect. The critical temperature of each unit is the $T_c$ of the corresponding device measured in [10].

The transition unit $I_{c0}$, on the other hand, is assumed to be the same for all TESs. Experimental data suggest that the critical current at zero temperature for a TES is roughly proportional to the spacing between the normal metal bars [10]. This is likely because in the superconducting state, the current meanders around the bars, and the width of the current path is the bar spacing. For devices with the same bar spacing, the $I_{c0}$ variation is small; therefore, we chose one TES design to obtain the unit $I_{c0}$ value: the 1/2-sq. 2-bar device with a total $I_{c0}$ of 7.8 mA and 5 inter-bar units gives a per-unit $I_{c0}$ = 1.48 mA. This value is applied to all other simulated devices.

According to Eq. (5a), the value of $c_R$ should be unity when the TES is in the normal state, to obtain the correct total normal state resistance, and $c_I$ should be unity when $T = 0$. However, they have both been experimentally observed to have a smaller value when the TES is biased in the transition [7][10], and theoretical considerations based on the phase-slip model of the superconducting transition also suggest $c_R$ should vary with temperature [6][11]. However, without any detailed model of the dynamics of phase-slip lines in 2-D films, or experimental measurements of the parameters those models might require, we kept $c_R$ and $c_I$ as fixed parameters in each of our simulations. Instead, we simply carried out separate calculations with manually chosen values of $c_I$ = 0.79 (as measured in [10]) and $c_R$ = 0.25, 0.5, and 1, to account for its variation. Any one calculation presented here is only strictly valid, therefore, for the region of the transition where $c_I$ and $c_R$ are close to these chosen values. Until a model for these two-fluid parameters is developed and we can identify those regions, we present our results over the full transition width.

## 3.2. Simulation results

Figure 5 shows the simulated resistance and current distribution of the 1-sq. 3-bar devices under different biases. Operating the device at a fixed temperature of $T = 72$ mK, as the TES bias current $I_{bias}$ increases from zero, several phenomena develop in the simulation. Initially the current flows through the device encountering no resistance (or very little, due to the approximation described previously – Fig. 3). This is because, although there are normal metal (resistive) regions present, a lower resistance path that meanders around the bars is present. Further increasing $I_{bias}$ will cause the current through a unit to increase, along with the total resistance along the meander path. Under this scenario, the current will still meander around the bars, because the total resistance of that long path is still below the equivalent resistance that the current would experience if it would go via the shorter, direct path intersecting the bars. Fig. 5(a) and Fig. 5(b) illustrate this for a device biased at ~ 12% $R_n^{total}$. Further increasing $I_{bias}$ will increase the transition unit resistance to a point where the direct path now has lower resistance, and so it is preferred by the current. This is illustrated in Fig. 5(c) and Fig. 5(d), at ~ 44% $R_n^{total}$. This behavior indicates that the current flow pattern in a TES with bars is dependent on the bias position in the transition. Similar current distribution dependence on bias has been reported in Ref. [9], although obtained via a different modelling and measurement technique.

Repeating this calculation at different values of $T$ generates a 3D map of the $R(I, T)$ function for this device, as shown in Fig. 5(a) for a 2-sq. 3-bar device, and Fig.5(b) for a 2-sq. 5-bar device. Comparing the two $R(I, T)$ surfaces shows (Fig. 6) that when biased beyond ~ 20% $R_n^{total}$, the transition shapes of the two devices are about the same. This is because although the two devices have a different number of bars and different bar spacing, at those biases the current is flowing through the TES uniformly, irrespective of the bar layout. Conversely, while biased below ~ 20% $R_n^{total}$, the device with the larger bar spacing (2-sq. 3-bar) shows a sharper transition; this is due to the wider current path available between bars. It can therefore support a larger critical current, making the transition width narrower. This difference is more evident in the comparison of the thermal sensitivity $\alpha$, and current sensitivity $\beta$, defined as:

$$\alpha = \frac{\partial logR}{\partial logT}\bigg|_{(I_0,T_0)}, \quad \beta = \frac{\partial logR}{\partial logI}\bigg|_{(I_0,T_0)}, \quad (6)$$

shown in Fig. 6(c) and Fig. 6(d) respectively. $(I_0, T_0)$ are the TES bias points represented by the black trajectories on the respective $R(I, T)$ surfaces in Fig. 6(a)(b). These trajectories have been obtained by combining the simulated $R(I, T)$ with the relation for the power flowing from the TES to the bath, assuming no incident photons:

$$I^2 R = G(T - T_{bath}), \quad (7)$$

where $G = 100$ pW/K is the thermal conductance between the TES and the thermal bath, and $T_{bath} = 55$ mK is the bath temperature. These curves represent the experimentally measurable I-V relation under voltage bias. Both devices exhibit similar trends for $\alpha$ and $\beta$ as a function of % $R_n^{total}$. Initially the values are fairly small, followed by a sudden increase between 10% and 20% $R_n^{total}$, corresponding to the steep region in the $R(I, T)$ surface. In this region, as the bias increases, more transition units change from the superconductive state to the steep resistive state as represented in Fig. 3. $\alpha$ and $\beta$ for the 2-sq. 3-bar device reach values almost double those reached by the 2-sq. 5-bar device, representative of the enhanced steepness of its transition. This is a well-known phenomenon associated with the presence of normal metal bars on TESs – increasing the number of normal metal bars reduces $\alpha$. Biasing the device higher in the transition results in a reduction of both quantities, due to the decrease in steepness of the resistive transition in both $I$ and $T$. This is also a phenomenon reported in many experimental papers, for example Ref. [2]. At even higher bias (> 40% $R_n^{total}$) the curves for the two devices collapse onto each other, due to the redistribution of the current in the network. The $R(I, T)$ surface is no longer affected by the presence of the bars, but only by the lateral dimensions of the TES, identical in both devices.

Following the same approach, and maintaining the values of $G$ and $T_{bath}$ fixed, we have run simulations for all the seven type of devices described above. Fig. 7 compares $\alpha$ for these devices. As in the previous discussion, $\alpha$ scales inversely with the number of bars for otherwise identical devices, independently of the device aspect ratio. Also, devices with an increasing aspect ratio, but same number of bars, show an increase in $\alpha$ over a wide range of % $R_n^{total}$ biases. Both these results agree with what is experimentally seen in [10], suggesting this model is correctly calculating the changes in TES transition steepness due to either the device dimensions (i.e. $R_n^{total}$ and consequently $I_0$) and/or the number and spacing of bars (i.e. $I_{c0}$)[10]. These results demonstrate the predictive power of this model.

The previously described simulations all use a fixed value for $c_R = 0.5$, chosen as an average value at low bias among those reported for these kinds of devices in [10]. However, $c_R$ varies through the transition, and it is smaller at lower bias points and increases with bias. Eq. (5a) shows that a smaller $c_R$ will result in a lower resistance in the transition region units, and consequently the current will prefer to meander around the normal metal bars until higher biases, and vice versa for higher values of $c_R$. Varying $c_R$ can consequently generate a family of curves of the kind showed in Fig. 8 for a given device. Due to the lack of experimental data on the dependence of $c_R$ on the bias and the difficulty in estimating the materials parameters that determine this phenomenological parameter (for ex-ample, the charge imbalance relaxation length [7]), for now its influence on TES transition shape can only be evaluated qualitatively. Figure 8 shows the values of $\alpha$ throughout the transition for the 1-sq. 3-bar device when using $c_R = 1$, 0.5, and 0.25. Based on the $c_R$ values measured in Ref. [10], this plot is divided into 3 regions; each region provides α values with a $c_R$ best estimated at that bias. The actual dependence of $\alpha$ on % $R_n^{total}$, i.e. the actual shape of the TES transition, can likely be approximated by a combination of these three curves. The model will be fairly straightforwardly improved if a quantitative model for $c_R$ and $c_I$ can be developed.

## 4. Conclusions

While the engineering of TES transition shapes has been performed experimentally through the manipulation of 2-D features, a predictive 2-D model has until now been lacking. In this paper, we present a 2-D resistor network model which can calculate the current distribution and overall $R(I, T)$ surface for TES devices with arbitrary geometry including normal metal features. The TES is divided into 4-terminal units, with resistances based on the superconducting two-fluid model and calculated self-consistently based on the temperature and net current, allowing for the calculation of the current distribution and total resistance of the device. The model has been used to simulate the transition shape and the current-voltage characteristics of a series of previously experimentally measured devices of varying dimensions and normal metal features. The simulations show how the normal metal features force the current flowing through the TES to meander around them at lower biases, while at higher biases the current tends to flow more uniformly through the entire width of the device, independently of the specific bar arrangement. The simulations also show how the different bar designs affect the steepness of the TES transition, replicating phenomena experimentally observed, such as the dependence of $\alpha$ on the number of bars. Our model shows qualitative agreement with experimental results, and therefore represents a powerful tool to guide the design of TESs with normal metal features and other non-uniform geometries. In the future, more complex effects such as localized heating and the lateral proximity effect could be implemented in this 2-D model. It may also be possible to study the noise mitigating mechanisms of the normal metal features.

## Acknowledgements

This work was supported by the Accelerator and Detector R&D program in Basic Energy Sciences' Scientific User Facilities Division at the Department of Energy and the Laboratory Directed Research and Development (LDRD) program at Argonne. This research used resources of the Advanced Photon Source and Center for Nanoscale Materials, U.S. Department of Energy Office of Science User Facilities operated for the DOE Office of Science by the Argonne National Laboratory under Contract No. DE-AC02- 06CH11357. The authors thank Kelsey M. Morgan for the very helpful discussion and suggestions.

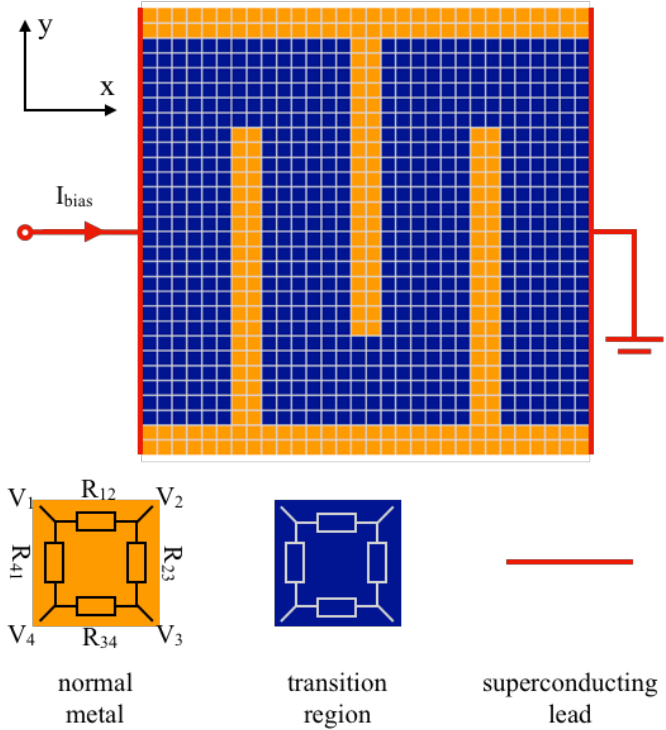

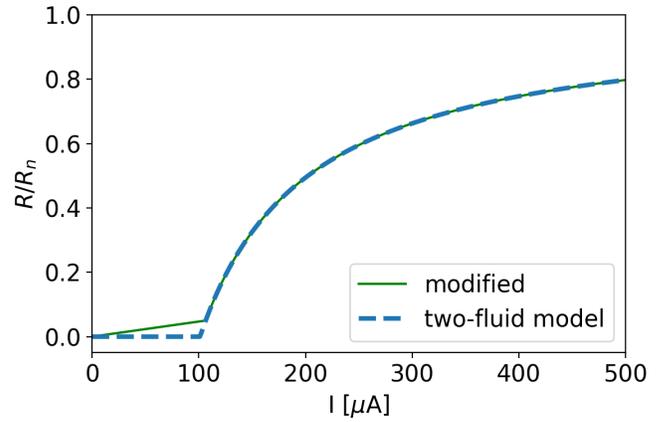

Figure 3: The two-fluid model $R$-$I$ relation (blue dashed line), and the modified $R$-$I$ relation (green line) that does not have zero-derivative region below 5% $R_n$ which is suitable for numerical iteration purpose.

Figure 1: The TES is divided into small units that are connected in a 2-D network. The basic unit is a 4-terminal resistor. Adjacent voltage nodes $V_1$, $V_2$, $V_3$, $V_4$ are electrically connected via resistances $R_{12}$, $R_{23}$, $R_{34}$, and $R_{41}$, respectively. $R_{12} = R_{23} = R_{34} = R_{41} = R$, and the total current is defined by $I = (I_x^2 + I_y^2)^{1/2}$, where $x$ and $y$ are longitudinal and transverse directions. The 2-D resistor network model of a 150 μm × 150 μm TES has normal metal banks (orange) on the edge along the current flow direction, three normal metal bars (orange) perpendicular to the current flow, superconducting leads (red), and the transition region (blue) where the resistances are a function of current and temperature. Each unit is a 5 μm × 5 μm square.

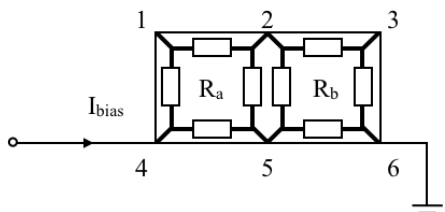

Figure 2: A simple 1×2 network has 2×3 = 6 nodes. The resistances in the first unit are denoted by $R_a$, and those in the second unit are $R_b$. A bias current $I_{bias}$ is applied to node-4. Node-6 is grounded and its voltage is $V_6 = 0$.

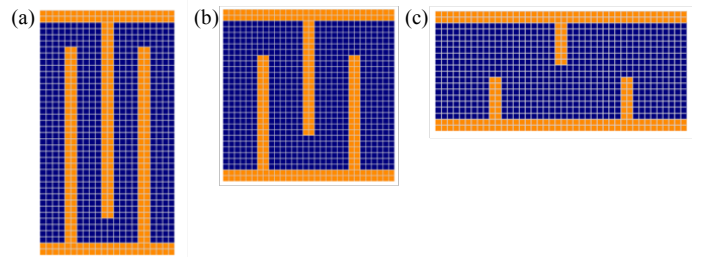

Figure 4: The 2-D layout of the TESs under study: (a) 106 μm × 212 μm (1/2-sq.) TESs are made with 3 bars and 2 bars; (b) 150 μm × 150 μm (1-sq.) TESs are made with 4 bars and 3 bars; (c) 212 μm × 106 μm (2-sq.) TESs are made with 5 bars, 4 bars, and 3 bars.

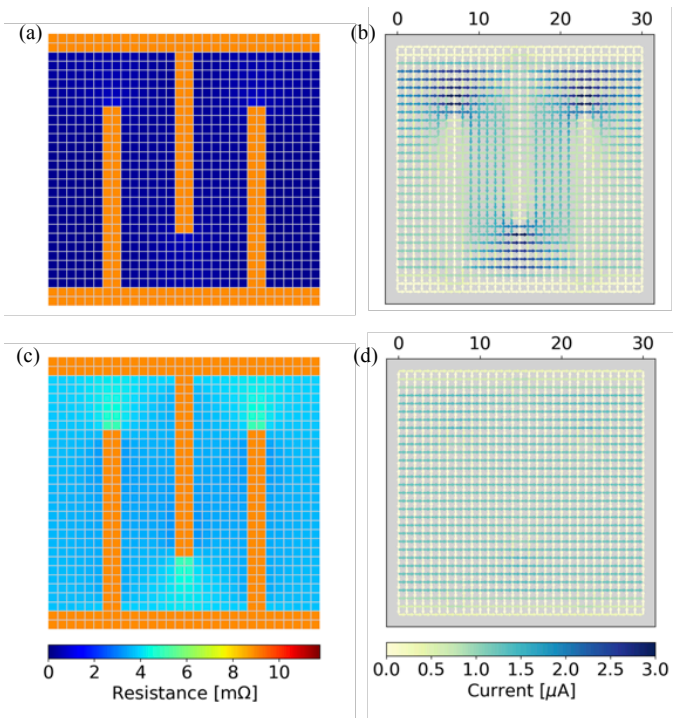

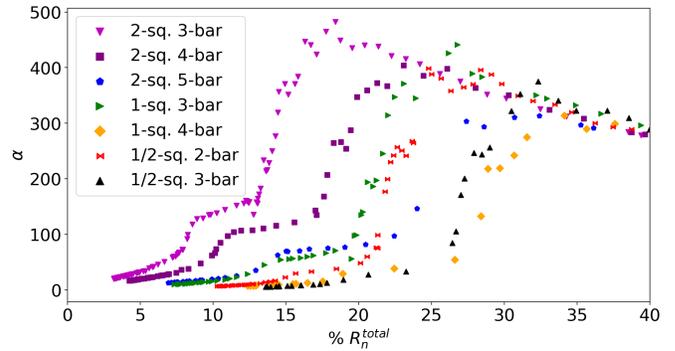

Figure 5: The resistance and current distribution of the 1-sq. 3-bar TES under different bias conditions. (a) (b) $I_{bias} = 35$ μA, T = 72 mK, and the TES at 12% $R_n^{total}$. (c) (d) $I_{bias} = 35$ μA, T = 72.8 mK, and the TES at 44% $R_n^{total}$.

Figure 7: The α values throughout the transition for the seven TESs. They show a clear correlation with the device design details, such as number of bars, spacing between bars and number of squares in the low bias regime.

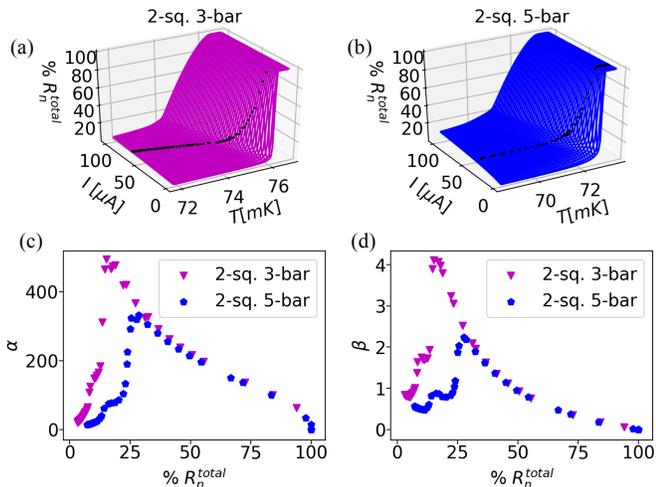

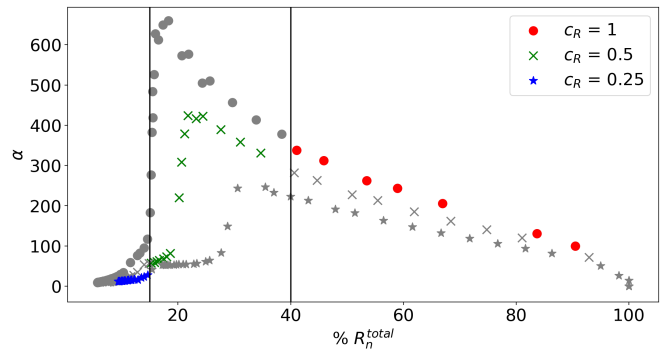

Figure 6: TESs with the same shape but different bar spacing: (a) 3-bar, (b) 5-bar. The transition shape changes around 20% $R_n^{total}$ due to the current distribution pattern change. At low bias, the transition of the 3-bar TES is sharper due to its larger bar spacing, which is clearly shown by the α and β values in (c) and (d), respectively.

Figure 8: The α values of the 1-square 3-bar TES under assumptions of $c_R = 1$ (round dots), $c_R = 0.5$ (crosses), and $c_R = 0.25$ (stars). The plot is divided into three regions by the 15% and 40% boundary. The α values that are using $c_R$ values suitable for each region are colored, while others are marked in grey. The actual dependence of α on % $R_n^{total}$ is a combination of the entire set of curves that can be generated by varying $c_R$ from 0 to 1, of which the three present are a representative example.